\documentclass[fleqn,10pt]{wlscirep}
\usepackage[switch]{lineno}

\PassOptionsToPackage{english}{babel}
\usepackage[normalem]{ulem}
\usepackage{times}
\usepackage{graphicx}
\usepackage{subcaption}
\usepackage{ragged2e}
\usepackage[english]{babel} 
\usepackage{siunitx}
\usepackage[utf8]{inputenc}
\usepackage{textcomp}
\usepackage{upgreek}
\usepackage{amsmath}

\DeclareCaptionJustification{justified}{\justifying}
\begin{document} 

\author[1]{\mbox{K. Baumgärtner}}
\author[2]{\mbox{M. Reuner}}
\author[1]{\mbox{C. Metzger}}
\author[3]{\mbox{D. Kutnyakhov}}
\author[3]{\mbox{M. Heber}}
\author[3]{\mbox{F. Pressacco}}
\author[1,4]{\mbox{C.H. Min}}
\author[1,3]{\mbox{T.R.F. Peixoto}}
\author[5]{\mbox{M. Reiser}}
\author[5]{\mbox{C. Kim}}
\author[5]{\mbox{W. Lu}}
\author[5]{\mbox{R. Shayduk}}
\author[5]{\mbox{W. M. Izquierdo}}
\author[3]{\mbox{G. Brenner}}
\author[6]{\mbox{F. Roth}}
\author[1]{\mbox{A. Schöll}}
\author[5,6]{\mbox{S. Molodtsov}}
\author[3,7,8]{\mbox{W. Wurth$\dagger$}}
\author[1]{\mbox{F. Reinert}}
\author[5]{\mbox{A. Madsen}}
\author[2,8]{\mbox{D. Popova-Gorelova}}
\author[3,5*]{\mbox{M. Scholz}}
\affil[1]{Experimentelle Physik 7, Julius-Maximilians-Universität, Am Hubland, 97074 Würzburg, Germany}
\affil[2]{I. Institute for Theoretical Physics and Centre for Free-Electron Laser Science, Universität Hamburg, Luruper Chaussee 149, 22607 Hamburg, Germany}
\affil[3]{Deutsches Elektronen-Synchrotron DESY, Notkestrasse 85, 22607 Hamburg, Germany}
\affil[4]{Institut für Experimentelle und Angewandte Physik, Christian-Albrechts-Universität zu Kiel, 24098 Kiel, Germany}
\affil[5]{European X-Ray Free-Electron Laser Facility, Holzkoppel 4, 22869 Schenefeld, Germany}
\affil[6]{Institute of Experimental Physics, TU Bergakademie Freiberg, Leipziger Strasse 23, 09599 Freiberg, Germany}
\affil[7]{Institut für Experimentalphysik, Universität Hamburg, Luruper Chausee 149, 22761 Hamburg, Germany}
\affil[8]{The Hamburg Centre for Ultrafast Imaging (CUI), Luruper Chaussee 149, 22607 Hamburg, Germany}
\affil[*]{To whom correspondence should be addressed; markus.scholz@desy.de.}

\title{Ultrafast molecular orbital tomography of a pentacene thin film using time-resolved momentum microscopy at a free-electron laser}


\begin{abstract}
We use time-resolved momentum microscopy at a free-electron laser (FEL) and extend orbital tomography into the time domain to image the electronic wave functions of excited molecular orbitals. This technique provides unprecedented insight into the ultrafast interplay between structural and electronic dynamics. In this work we prove general applicability and establish the experimental conditions at FEL sources to minimize space charge effects and radiation damage. We investigate a bilayer pentacene film on Ag(110) by optical laser pump and FEL probe experiments. From the momentum microscopy signal, we obtain time-dependent momentum maps of the excited-state dynamics of both pentacene layers separately. Combining experimental observations with a theoretical study, we interpret the observed signal for the bottom layer as resulting from the charge redistribution between the molecule and the substrate induced by excitation. We identify that the dynamics of the top pentacene layer resembles excited-state molecular dynamics.  
\end{abstract}

\maketitle 
\section*{Introduction}

Photo-induced phenomena of adsorbates on solid surfaces have been intensively investigated over the last decade \cite{dantus_realtime_1987,zewail_laser_1988,petek_real-time_2000,frischkorn_femtochemistry_2006}. Despite some success in controlling physical properties and initiating femtochemistry by light, the description of ultrafast dynamics upon excitation remains challenging. Significant advances in photo-induced surface chemistry require deciphering the complex interplay between excited electronic wave packet dynamics as well as the rapid rearrangement of atomic positions and interactions at the metal-organic interface. Angle-resolved photoelectron spectroscopy (ARPES) is a well-known and powerful method to investigate the electronic structure of molecules. In the last decade, orbital tomography has emerged as an exciting extension of the photoemission technique for imaging localized electronic wave functions in thin film molecules \cite{ziroff_hybridization_2010-1,puschnig_reconstruction_2009,dauth_orbital_2011,weis_exploring_2015,graus_electron-vibration_2016,YangJPhChLett19}. In this framework, the photoemission process can be described either in a one-step model or using more sophisticated final state approximations \cite{luftner_understanding_2017,grimm_molecular_2018,dauth_perpendicular_2016,metzger_plane-wave_2020, Popova-GorelovaPRA16}. Although the phase of the electronic wave function is not an observable, it can be retrieved under suitable experimental conditions \cite{wiesner_complete_2014}, or by use of iterative algorithms traditionally employed in coherent diffraction imaging \cite{luftner_imaging_2014,kliuiev_algorithms_2018}. This enables more intricate data analysis, like reconstructing of the electron density of molecular orbitals in real space.\\

Expanding the orbital tomography technique into the time domain requires XUV or x-ray photon energies and ultrashort pulses with sufficient flux presently only provided by self-amplified spontaneous emission (SASE) free-electron lasers (FELs) or high harmonics generation (HHG) sources. FELs based on superconducting linear accelerators such as FLASH at DESY in Hamburg \cite{tiedtke_soft_2009} offer not only tunable photon energies but also repetition rates ranging from hundreds of kHz to MHz. In the near future full polarization control of the x-ray pulse with a duration of $<$ 10\,fs with sufficient fluence will be available \cite{rohlsberger_light_2019,doi:10.1063/5.0018834}. FELs are thus an excellent source for a variety of pump-probe experiments such as time-resolved orbital tomography. However, SASE pulses have strong shot-to-shot intensity fluctuation and the data must be sorted by temporal jitter and pulse energy accordingly. The high brightness of a FEL can lead to radiation-induced damage in the sample and unwanted space charge effects in photoelectron experiments that limit the energy and momentum resolution in time- and angle-resolved photoemission spectroscopy \cite{dellangela_vacuum_2015}. For soft matter, in particular, the influence of radiation damage may result in significant peak broadening or side features in the spectra caused by bond-breaking or radical formation \cite{amy_radiation_2006}. A careful survey of the sample degradation under illumination at various light intensities and time scales is therefore crucial to gauge and suppress radiation damage. The effects of optical pump and x-ray probe pulse-induced space charge in photoemission were intensely studied over the last years, both from a theoretical point of view and by experiments \cite{hellmann_vacuum_2012,dellangela_vacuum_2015,schonhense_correction_2015,schonhense_multidimensional_2018}. Coulomb interaction between excited photoelectrons during and after the FEL pulse can lead to a significant shift in binding energy, spectral broadening, and a smeared out photoelectron angular distribution. Pump laser multiphoton excitation results in a cloud of slow electrons in the vicinity of the sample surface and perpendicular to the axis of the flight path in the time-of-flight (TOF) instrument. On their trajectories towards the detector they can interact with faster photoelectrons passing by and cause an additional shift and broadening in the detected photoemission signal. It is thus essential to reduce the number of slow electrons and photoelectrons per bunch by attenuating the optical laser and FEL \cite{hellmann_vacuum_2012,dellangela_vacuum_2015}.\\

Studying the formation and relaxation dynamics of charge carriers and electron transfer between molecules and at the molecule-substrate interface are of crucial importance for the optimization of organic device properties. Pentacene is a well-characterized molecule which has attracted considerable interest in the scientific community due to its its potential to exceed the Shockley-Queisser limit in solar energy conversion \cite{congreve_external_2013,zhu_charge-transfer_2009}. Specifically, we investigate a bilayer of pentacene atop Ag(110). This system exhibits a well-ordered growth structure with high reproducibility. As recently shown \cite{grimm_molecular_2018}, the spectral features in the electronic structure of the first and second layer are well separated in energy from each other. While in the bottom layer the lowest unoccupied molecular orbital (LUMO) is partly filled due to charge transfer from the substrate, the top layer is largely electronically decoupled. This simplifies the identification and reconstruction of the orbitals of individual layers and the capturing of ultrafast charge transfer dynamics within frontier orbitals and at the molecule-substrate interface on a femtosecond time scale. 



\section*{Method}

The experiments were performed at the PG2 beamline at FLASH \cite{martins_monochromator_2006}. The experimental geometry is depicted in Fig.\,1\,(a). Optical pump and FEL probe pulses (both p-polarized) impinge on the sample at a polar angle of $\uptheta$\,=\,68$^{\circ}$ and an azimuthal angle of $\upphi$\,=\,64$^{\circ}$ with respect to the [-1\,1\,0] direction of the Ag(110) surface, and are aligned to have spatial overlap. The photoelectrons emitted into the hemisphere above the sample are detected by a novel TOF momentum microscope \cite{schonhense_space-_2015,kutnyakhov_time-_2020} and recorded according to their wave vector and kinetic energy. The photon wavelength of the FEL is 35\,nm and the average FEL pulse energy of 30\,$\mu$J is attenuated by nitrogen gas and thin film filter foils to acquire spectra similar to reference measurements in static experiments \cite{grimm_molecular_2018}. The energy resolution in the experiment is 80\,meV (see supplement, Fig.\,1\,(b)). The optical pump laser provides a maximum flux of 1\,mJ/cm$^2$ at 400\,nm wavelength, and is synchronized with the FEL to allow for pump-probe delay scans. The temporal cross-correlation between optical laser pump pulse and FEL probe pulse is $215\pm10$\,fs FWHM. Within a so-called bunch train of 330 pulses with 1\,$\mu$s spacing, at a repetition rate of 10\,Hz, 299 pulses are optically pumped, while the remaining 31 are unpumped (effective repetition rate of 3.3\,kHz). The temporal overlap of FEL and optical laser ($\uptau_{\mathrm{Delay}}=0$) is established by analyzing the intensity evolution at the Fermi edge of a bilayer of pentacene on Ag(110). The beam spot size of the FEL and optical laser at the sample position is about 250\,$\mu$m\,x\,150\,$\mu$m and 260\,$\mu$m\,x\,150\,$\mu$m, respectively. For capturing the dynamics, the pump pulses are synchronized with the FEL in such a way that the optical delay can be chosen freely between several tens of femtoseconds to tens of picoseconds with respect to the FEL bunch. A typical time-resolved orbital tomography measurement sampling over 3\,ps is obtained within $\sim$6\,h. Pentacene (purity 99\,\%, purchased from Sigma-Aldrich) is deposited from a homemade Knudsen cell evaporator at a deposition rate of one monolayer per 30\,min. The overall preparation time of the sample including annealing and sputtering cycles of the metal substrate is about 3.5\,h. Before conducting the photoemission experiment the film thickness and quality were verified by low-energy electron diffraction (LEED), evidencing the pentacene bilayer structure on Ag(110) \cite{grimm_molecular_2018} (see supplement, Fig.\,1\,(a)). In this superstructure, all molecules are aligned with the long molecular axis along the [001] direction of the silver substrate and tilted along their long axis by 6$^{\circ}$ and 8.5$^{\circ}$ within the first and second layer, respectively. The sample is kept at room temperature during the deposition and experiment. The base pressure in the analyzer chamber is $\sim1\cdot 10^{-10}$\,mbar.

\section*{Results and Discussion}

\begin{figure}[!h]
	\includegraphics[width=\linewidth]{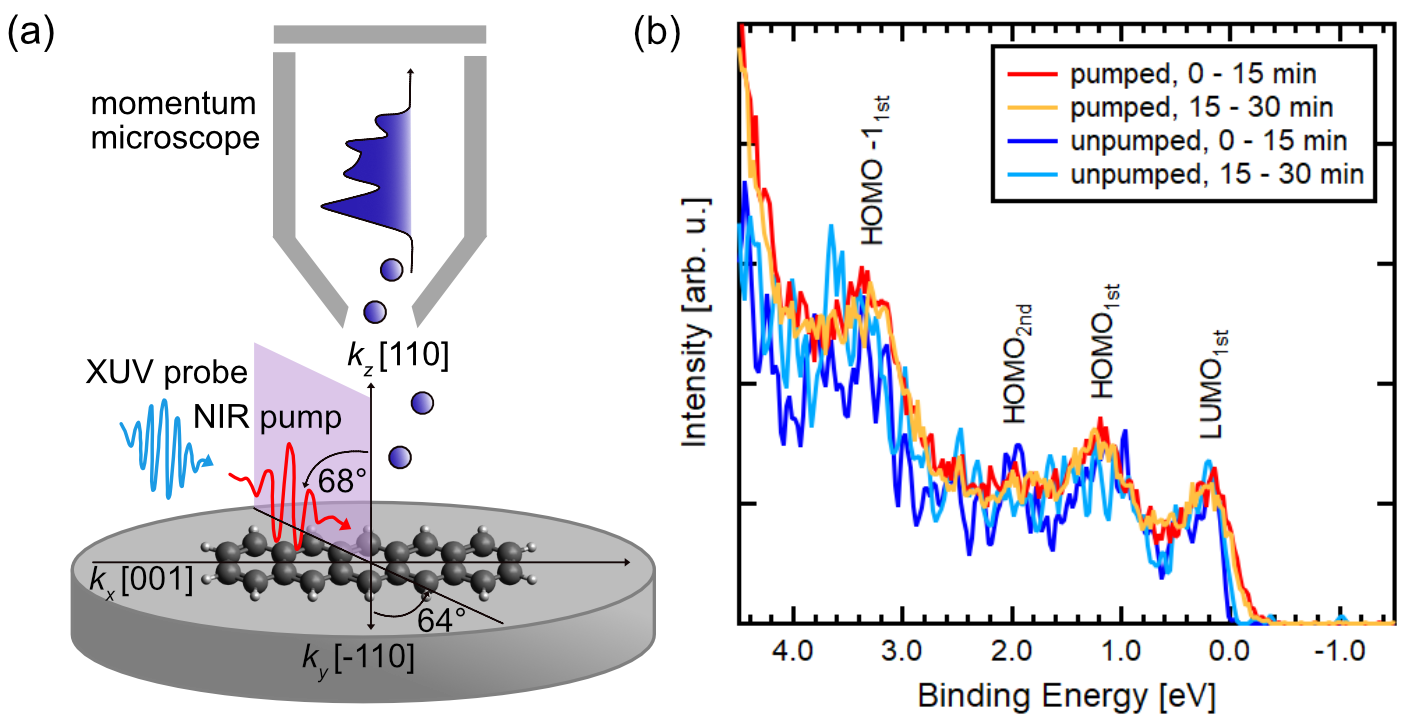}
	\caption{(a) Schematic illustration of the experimental geometry. All of the pentacene molecules are adsorbed with the long molecular axis along the [0\,0\,1] direction of the silver substrate. Pump and probe pulses are aligned to coincide in the same sample region. (b) Time-integrated photoelectron spectra for a bilayer of pentacene on Ag(110) with an excitation energy of h$\upnu$\,=\,35\,eV. The measurements are taken at the same sample position for a total measurement time of 30\,min. Within a bunch train of 330 pulses with 1\,$\mu$s spacing at a repetition rate of 10\,Hz, 299 bunches are optically pumped (red and orange lines). The dark and light blue lines show the time integrated signals of the 31 unpumped bunches. LUMO$_{\mathrm{1st}}$, HOMO$_{\mathrm{1st}}$, HOMO-1$_{\mathrm{1st}}$ of the first layer and HOMO$_{\mathrm{2nd}}$ of the second layer are indicated in the spectra. The spectra are normalized at 2.15\,eV binding energy. The LUMO$_{\mathrm{1st}}$ is partly filled due to charge transfer from the substrate.}
	\label{img:radiation}
\end{figure}

Fig.~\ref{img:radiation}(b) displays the normalized, angle-integrated energy distribution curve (EDC) of the frontier orbitals measured with 35\,eV photon energy, during constant illumination at the same sample position. The red curve shows the pumped \mbox{photoemission} signal integrated over the first 15\,min after the beginning of the illumination together with the following 15\,min of measurement (orange) and the parallel detection of the unpumped photoemission signals (dark and light blue). 
Several molecular features as well as the onset of the Ag 4d-bands towards higher binding energies are visible in the recorded valence region: the partly filled LUMO of the first layer (LUMO$_\textrm{1st}$), the highest occupied molecular orbital (HOMO) of the first layer (HOMO$_\textrm{1st}$), the HOMO of the second layer (HOMO$_\textrm{2nd}$), and the second highest occupied molecular orbital (HOMO-1) of the first layer (HOMO-1$_{\mathrm{1st}}$). Even though molecular thin films tend to be susceptible to temperature- and light-induced deterioration \cite{amy_radiation_2006}, there is no observable radiation damage within the error of the photoemission intensity for a total measurement time of 30\,min at the same sample position reasoned by the similarity of the red and orange curves as well as of the light and dark blue curves in Fig.\,1\,(b). As expected for the optically pumped photoemission signal, the spectral features are only marginally broadened in energy due to additional space charge-induced effects from slow electrons. They are also shifted in binding energy by about 150\,meV with respect to the EDCs obtained from the 31 unpumped pulses (Fig.\,1\,(b), light and dark blue). Note, this energy shift is corrected in Fig.~\ref{img:radiation}(b). Nevertheless, they are still easily discernible in the EDC and can be assigned to their corresponding molecular initial states by comparison to theoretical photoemission calculations.\\

\begin{figure*}
	\includegraphics[width=\linewidth]{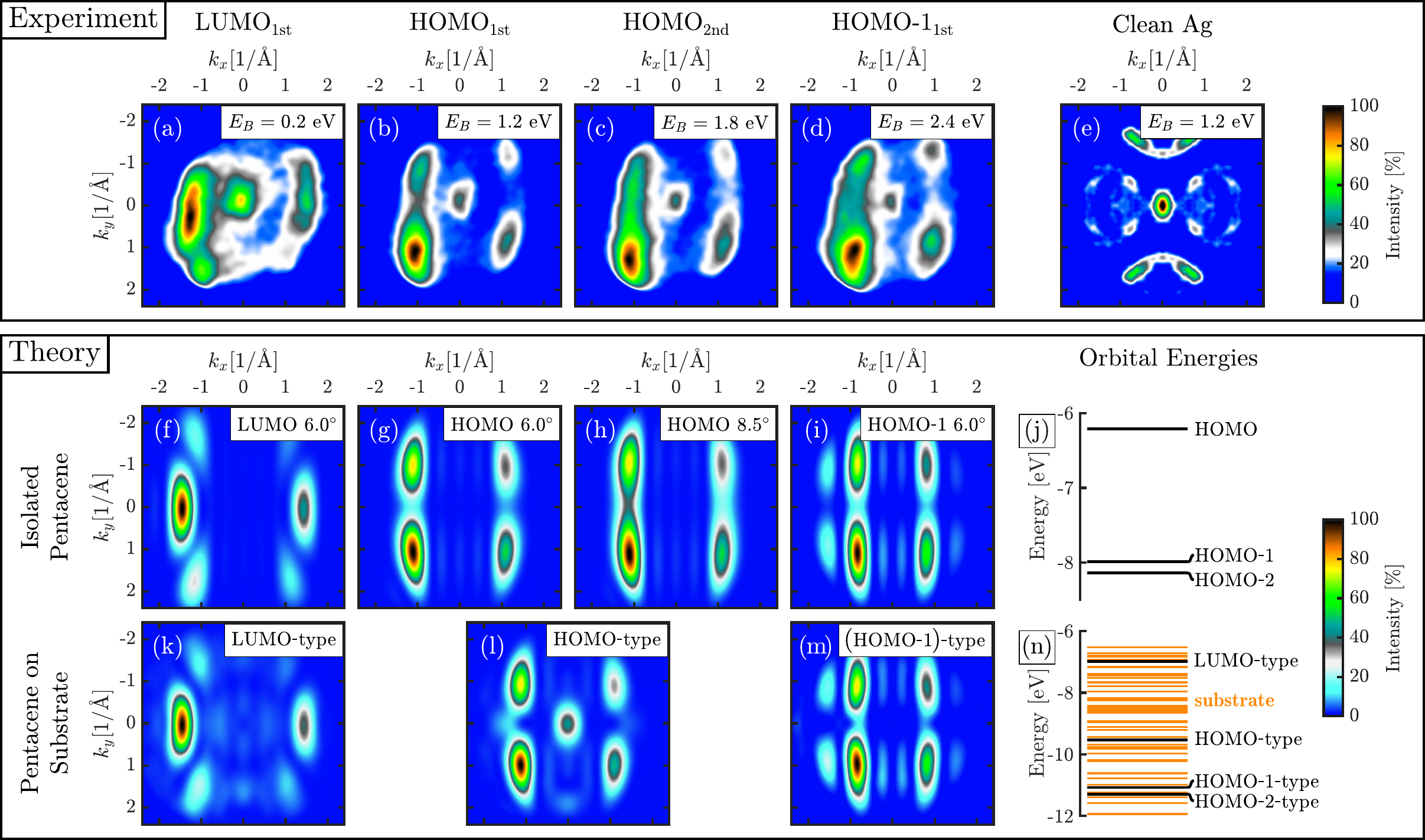}
	\caption{Time-integrated measured (a--d) photoelectron momentum maps (PMMs) for the pentacene valence orbitals at an excitation energy of $\mathrm{h}\upnu$\,=\,35\,\textrm{eV} under simultaneous illumination with the pump and probe pulses. The partly filled LUMO$_{\mathrm{1st}}$ (a), the HOMO$_{\mathrm{1st}}$ (b), HOMO$_{\mathrm{2nd}}$ (c), and the HOMO-1$_{\mathrm{1st}}$ (d) can be clearly distinguished from one another and identified by comparison to the simulations (f--i) and (k--m). The simulations take into account that the pentacene molecules are tilted along the long molecular axis by $6.0^{\circ}$ and $8.5^{\circ}$ in the first and second molecular layer, respectively \cite{grimm_molecular_2018}. (e) An unpumped, symmetrized PMM of clean Ag(110) at $\mathrm{E_B}$\,=\,1.2\,eV with otherwise identical parameters for comparison. The color scale in the experimental PMMs has been adjusted to suppress the background signal. Simulated PMMs for the (f) LUMO, (g) and (h) HOMO, and (i) HOMO-1 orbitals and (j) Hartree-Fock energies of occupied molecular orbitals of isolated pentacene. The alignment of isolated pentacene was adjusted to fit the experiment. (j--l) Simulated PMMs for the pentacene on Ag(110) cluster. The orbitals of the molecular (k) LUMO-, (l) HOMO-, (m) (HOMO-1)-type character and all orbitals with energies within the 500\,meV range (experimental energy averaging range) around their energies contribute to the simulated signal. (n) Hartree-Fock energies of the pentacene on Ag(110) cluster. Thick black lines outline energies of orbitals with a predominantly molecular-type character. 
	} 
	\label{fig:stationaryData}
\end{figure*}

Fig.~\ref{fig:stationaryData} (a--d) displays the experimental time-integrated photoelectron momentum maps (PMMs) of the pentacene film.
All momentum maps are integrated in an energy window of 500\,$\textrm{meV}$ centered at the maximum intensity position of LUMO$_\textrm{1st}$, HOMO$_\textrm{1st}$, HOMO$_\textrm{2nd}$, and HOMO-1$_{\mathrm{1st}}$. These four states in the valence region can unambiguously be identified as orbitals of predominantly molecular character using ab initio calculations. The orbitals of isolated pentacene and pentacene on the Ag(110) cluster are calculated using Hartree-Fock theory. We use the Molcas software package \cite{https://doi.org/10.1002/jcc.24221_Molcas} for all electronic-structure calculations. The PMMs calculated within the plane-wave approximation to the photoelectron wave function. They are proportional to the Fourier transform of the Dyson orbital $\phi_D(\mathbf r)$ squared, $|\mathcal F[\phi_D(\mathbf r)]|^2$, where the Dyson orbital $\phi_D(\mathbf r)$ is the overlap function between the $N$-electron wave function of an electronic system and its $(N-1)$-electron wave function produced by the emission of an electron \cite{puschnig_reconstruction_2009}. If the state of a system before ionization is a neutral ground state, the Dyson orbital is well approximated by the molecular orbital, from which a photoelectron was detached. Thus, we substitute Hartree-Fock orbitals for the calculation of PMMs in an unexcited state and compare experimental and stationary theoretical results. We find that the simulated PMMs for LUMO, HOMO and HOMO-1 of isolated pentacene in Fig.~\ref{fig:stationaryData} (f)--(i) reproduce the main features of the experimental PMMs quite precisely, which allows us to unambiguously identify them as orbitals of predominantly molecular character. There is already a good agreement between the experimental and calculated results for an isolated pentacene, including the spectral weight asymmetry resulting from the experimental geometry and polarization of the probe pulse. While similar comparisons are routine in the evaluation of static photoemission data, they are unprecedented in a pump-probe scheme at a FEL.\\ 

\begin{figure}
    \centering
    \includegraphics[width=\textwidth]{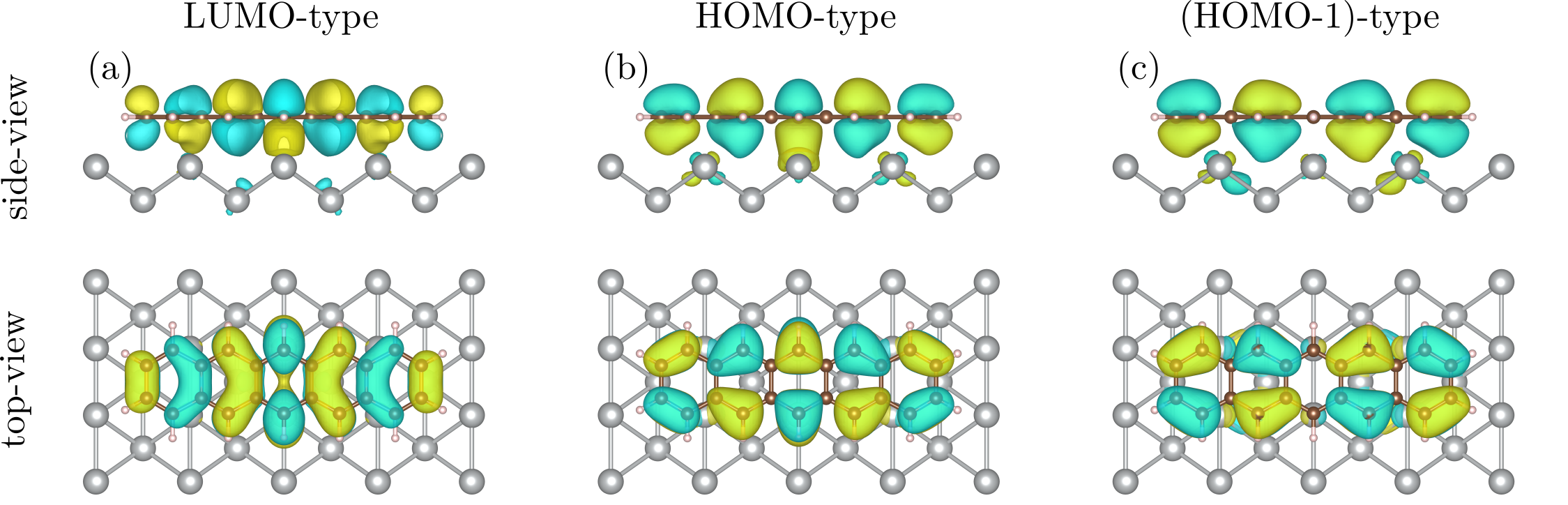}
    \caption{Calculated orbitals of pentacene on the Ag(110)-cluster with the major contribution of orbitals of molecular (a) LUMO-, (b) HOMO- and (c) (HOMO-1)-type character. The orbitals are visualized using the VESTA software \cite{MommaJAC11}.}
    \label{fig:pentaceneAgHybOrb}
\end{figure}

In addition to the features in the PMM due to molecular-type orbitals, the central feature close to normal emission is present in all experimental PMMs in Fig.~\ref{fig:stationaryData} (a--d). In order to explain this peak at the $\Gamma$-point and better understand the charge transfer between pentacene and the substrate, we calculated the electronic structure and PMMs of a model system consisting of a pentacene and a cluster of silver atoms (see SI for details). We obtain that the orbitals of this system are of three types: ones that have only silver contribution, ones that have a major contribution of molecular-type orbitals and some contribution of silver orbitals, and orbitals that have a minor contribution of molecular-type orbitals and a major contribution of silver orbitals. In our calculations, we reproduced the occupation of the LUMO-type orbital, which indicates that the molecule-substrate interaction is relatively strong and cannot be explained by van-der-Waals interactions \cite{abbasi2009ab_MP2Method}. This interaction leads to the emergence of hybridized bonds between the molecule and the substrate as can be observed in Fig.~\ref{fig:pentaceneAgHybOrb} showing orbitals of predominantly molecular character. The relative energies of the molecular-type orbitals hybridized with the substrate are slightly shifted in comparison to the isolated molecular orbitals (see Figs.~\ref{fig:stationaryData}(j) and (n)). For the simulation of the PMMs in Fig.~\ref{fig:stationaryData} (j--l), we took into account that all experimental momentum maps were integrated in an energy window of 500 meV. Thus, we included all orbitals with the energies in the range of 500 meV around the energy of the orbitals of predominantly molecular character in the calculation of PMMs. We reproduced the peak at the $\Gamma$-point for the HOMO-type orbital in Fig.~\ref{fig:stationaryData}(b) and (c), which arose due to the contribution of orbitals of all three types. We could not reproduce the central features for all PMMs likely because electrons from Ag atoms in the deeper layers also contribute to the experimental signal, enhancing substrate features in the PMMs (see Fig.~\ref{fig:stationaryData}(e)). This feature is also visible for clean Ag(110) and stems from silver sp-bands, see Fig.~\ref{fig:stationaryData}(e).

\begin{figure*}
	\includegraphics[width=\linewidth]{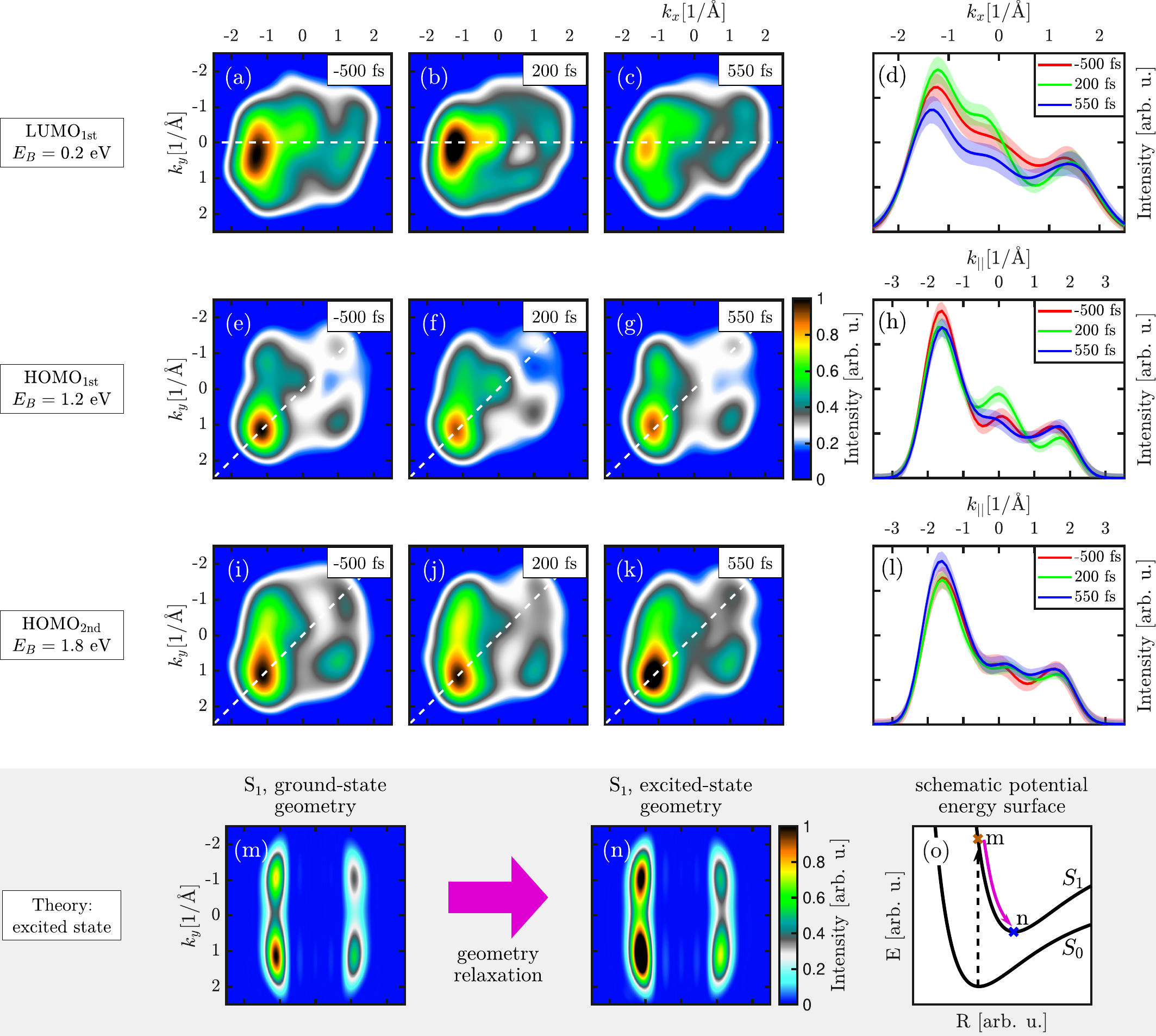}
	\caption{Time evolution of photoelectron momentum maps (PMMs) of pentacene of the (a)--(c) LUMO$_{\mathrm{1st}}$, (e)--(g) HOMO$_{\mathrm{1st}}$ and (i)--(k) HOMO$_\mathrm{2nd}$ at $\mathrm{E_B}$\,=\,0.2\,eV, $\mathrm{E_B}$\,=\,1.2\,eV and $\mathrm{E_B}$\,=\,1.8\,eV binding energy, respectively. The PMMs are integrated in intervals of 350\,fs and smoothed in momentum space with a Gaussian filter. Momentum distribution curves for (d) LUMO$_{\mathrm{1st}}$, (h) HOMO$_{\mathrm{1st}}$ and (l) HOMO$_{\mathrm{2nd}}$ (bottom) in the direction as indicated by the dashed lines in the corresponding PMMs. The shaded bands reflect 1\,$\sigma$ error-bars. Calculated PMMs for a pentacene molecule excited to the first singlet state (m) in the ground state geometry and (n) excited-state geometry. (o) Schematic representation of the proposed molecular dynamics in the second layer after excitation: the shape of a molecule changes after the electronic excitation.}
	\label{img:dyn_orbitals}
\end{figure*}

Besides these time-integrated data sets, the high repetition rate of the FEL and high efficiency of the momentum microscope enables resolving the time-dependent PMMs down to the femtosecond scale, as depicted in Fig.\,4\,(a) for the photoemission signal of LUMO$_\textrm{1st}$, HOMO$_\textrm{1st}$, HOMO$_\textrm{2nd}$ at $\mathrm{E_B}$\,=\,0.2\,$\textrm{eV}$, $\mathrm{E_B}$\,=\,1.2\,$\textrm{eV}$, and $\mathrm{E_B}$\,=\,1.8\,$\textrm{eV}$ binding energy, respectively. The PMMs have been integrated over 350\,$\textrm{fs}$ with the center at the indicated pump-probe delay $\uptau_{\mathrm{Delay}}$ and are smoothed in momentum space with a Gaussian filter with a FWHM of $0.16\,$\AA$^{-1}$. As already apparent for $\uptau_{\mathrm{Delay}}$\,$<$\,0 and in Fig.\,1\,(b), the LUMO is partially occupied due to charge transfer from the substrate. After optical excitation, the characteristic molecular features of LUMO$_\textrm{1st}$ and HOMO$_{\mathrm{1st}}$ show subtle changes. The intensity of the molecular features of HOMO$_{\mathrm{1st}}$ drop after time zero while the intensity of the partly occupied LUMO$_{\mathrm{1st}}$ slightly increases. The most significant intensity change is around the $\Gamma$-point. For both, LUMO$_{\mathrm{1st}}$ and HOMO$_{\mathrm{1st}}$, the intensity around the center increases after time zero and drops after $\uptau_{\mathrm{Delay}}$\,$=$\,200\,fs. The observed dynamics of the wavepacket can be explained in the framework of charge transfer between molecules and substrate \cite{xiang_ultrafast_2017}. After time zero, charge is redistributed from HOMO$_{\mathrm{1st}}$ to the partly filled LUMO$_{\mathrm{1st}}$, apparent in Fig.~\ref{img:dyn_orbitals}(b) by the variation of the total intensity of the molecular features. This process is most probably accompanied by the change of the molecular shape, since molecular equilibrium geometry strongly depends on the electronic state. Due to the strong interaction with the substrate, the molecule can also move relative to the surface after excitation, which is supported by the computational observation that the strength of the intensity of the central feature depends on the distance between the molecule and the substrate (see SI). While on clean Ag substrate, the lifetime of electronic excitation is typically a few femtoseconds \cite{bauer_dynamics_2002}, the observed dynamics around the $\Gamma$-point are about two orders of magnitude slower and correspond to the time scales of atomic motions in a molecule. We thus suggest that the dynamical response of the overall shape in the PMMs of LUMO$_\textrm{1st}$ and HOMO$_{\mathrm{1st}}$ and in the intensity around the $\Gamma$-point reflects the charge redistribution between the molecule and the substrate followed by a change in adsorption height and shape of the molecule.

In contrast, time-resolved PMMs in Fig.~\ref{img:dyn_orbitals}(i)--(k) show a nearly constant $\Gamma$-point feature at the HOMO of the second pentacene layer. This can be explained by a weak interaction of the second pentacene layer with the substrate and thus its dynamics could be closer to the behavior of isolated pentacene molecules upon excitation. In order to verify this hypothesis, we calculated PMMs of the excited isolated pentacene molecule. Since the experimental energies of the first and second bright singlet excited states, $S_1$ and $S_2$, are 2.3 eV and 3.7 eV, correspondingly \cite{HalasinskiJPhChA00}, a molecule should be most probably excited to $S_1$ by the excitation energy of 3.1 eV. The $N$-electron wave function in $S_1$ and $(N-1)$-electron wave functions of possible final states after the ionization that are needed for the calculation of the Dyson orbital are obtained using the RASSCF approach \cite{MalmqvistJPhCm90} (see SI for details). Upon excitation, the atoms of the molecule rearrange from the equilibrium geometry of the ground state to the equilibrium geometry of the excited state. Using the second order multiconfigurational perturbation theory \cite{AnderssonJPhCh90, AnderssonJCmPh92}, we find the optimized geometry of pentacene in $S_1$ and compare PMMs of excited pentacene in the ground-state geometry and in the excited-state geometry. We find that the shape change of pentacene in the singlet excited state appears as a broadening of the strongest peak in the PMM (the peak in the left bottom in Figs.~\ref{img:dyn_orbitals}(m)--(n)). This agrees very well with the peak broadening in the experimental data for the second pentacene layer in Fig.~\ref{img:dyn_orbitals}(k). We also find that the geometrical relaxation in $S_1$ leads to an overall intensity increase of the signal, which is also observed in the experimental data after the excitation (blue and green lines in Fig.~\ref{img:dyn_orbitals}(l)). The broadening and intensity increase are less pronounced in the experiment because both excited and unexcited pentacene contribute to experimental PMMs. The reported life time of the singlet excited state in pentacene dimers of about 0.5 ps \cite{ZirzlmeierPNAS15} allows us to assume that the singlet fission process has not started during the experimental time frame. This excellent agreement indicates possible perspectives for employing the second layer of the pentacene dimer absorbed on the substrate as an experimental model system simulating the electron dynamics of an aligned isolated pentacene molecule. Without a need to apply challenging techniques of molecular alignment in a gas-phase, such systems combined with our method of time-resolved orbital tomography can open up new opportunities to perform molecular movies \cite{KaramatskosNatComm19, BlagaNature12, NicholsonScience18, Popova-GorelovaPRA16}. 

In summary, we have demonstrated that sub-picosecond time-resolved orbital tomography of molecular thin films are feasible at SASE FELs. We established the experimental conditions to minimize space charge induced effects and radiation damage. This allowed us to identify and distinguish pumped molecular orbitals from substrate derived signatures of a pentacene bilayer/Ag(110) by comparison with theoretical calculations of both isolated pentacene and pentacene adsorbed on the silver substrate. From the time-resolved PMMs we observe shape changes within the molecular features after excitation which are accompanied by intensity redistributions of molecular and substrate features. Excited-state theory calculations connect the changes of the molecular features to a rearrangement of atoms in a molecule. The simultaneous time evolution of molecular and substrate features in momentum maps indicates rearrangements in geometry and adsorption height of molecules bound to the substrate during interfacial charge transfer upon excitation. Given the upcoming improvements in time resolution towards sub-femtoseconds and full polarization control at FEL light sources alongside with progress in the theoretical framework based on quantum electrodynamics \cite{popova-gorelova_imaging_2018}, time-resolved tomography of molecular wave functions during chemical reactions based on our work will provide unprecedented insight into photo-induced dynamics. 

\section*{Acknowledgement}
This  work  is  dedicated  to Prof. Dr. Wilfried  Wurth,  who  passed  away  on  May  8,  2019. We thank Andreas Oelsner for his technical assistance. We thank Holger Meyer and Sven Gieschen from the University of Hamburg for support of HEXTOF setup. The authors would like to thank FLASH for beamtime, and the staff of FLASH for their support during the experiment. We acknowledge financial support from the DFG through the W\"urzburg-Dresden Cluster of Excellence on Complexity and Topology in Quantum Matter -- \textit{ct.qmat} (EXC 2147, \mbox{project-id 39085490}), the excellence cluster "The Hamburg Centre for Ultrafast Imaging - Structure, Dynamics and Control of Matter at the Atomic Scale" of the Deutsche Forschungsgemeinschaft (DFG EXC 1074), the SFB 925 (project B2), and projects SCHO1260/4-2 and RE1469/12-2. M.R. and D. P.-G. acknowledge the support of the Volkswagen Foundation.

\section*{Contributions}
M.S. designed the experiment. M.S., K.B., M.R. and C.M. performed the data analysis. M.R. performed the theoretical study. M.S., K.B., C.H.M., T.R.F.P., M.R., C.K., W.L., W.M.I., F.R., D.K., M.H., F.P. and W.W. conducted the experiment. M.S., K.B., M.R., C.M. and D.P.-G. wrote the manuscript. M.S., K.B., M.R., C.M., C.H.M., T.R.F.P., C.K., W.M.I., F.R., D.K., M.H., F.P., M.R., W.L., R.S., G.B., A.S., S.M., F.R., W.W., A.M. and D.P.-G. contributed to scientific discussions. 

\section*{Data availability}
The data that support the findings of this study are available from the first author upon request.

\bibliographystyle{naturemag}
\bibliography{FLASH_paper_new}{}


\end{document}